# Simultaneous Control and Human Feedback in the Training of a Robotic Agent with Actor-Critic Reinforcement Learning

**Kory Mathewson** and **Patrick M. Pilarski**
Departments of Computing Science and Medicine
University of Alberta, Edmonton, Alberta, Canada
[korym, pilarski] @ ualberta.ca

## Abstract

This paper contributes a preliminary report on the advantages and disadvantages of incorporating simultaneous human control and feedback signals in the training of a reinforcement learning robotic agent. While robotic human-machine interfaces have become increasingly complex in both form and function, control remains challenging for users. This has resulted in an increasing gap between user control approaches and the number of robotic actuators which can be controlled. One way to address this gap is to shift autonomy to the robot. Semi-autonomous actions of the robotic agent can then be shaped and refined by human feedback, simplifying user control. Most prior work on human-agent shaping has incorporated training with feedback, or has included indirect control signals. By contrast, in this paper we explore how a human can provide concurrent feedback signals and real-time myoelectric control signals to train a robot's actor-critic reinforcement learning control system. Using both a physical and a simulated robotic system, we compare training performance on a simple movement task when reward is derived from the environment, from the human, and from the combination of the two. Our results indicate that benefit may be gained by including human generated feedback in learning algorithms for this complex human-machine interactive domain.

## 1 Introduction

Human-robot interaction is becoming more complex with the advancement of actuator and sensor technologies. One key example is that of robotic prostheses: artificial limbs attached to the body to replace abilities lost through injury or illness. Prosthetic limbs that have comparable degrees of freedom (DoF) and movement to human limbs have now been developed. A principal limitation is the complex control of such devices [Castellini *et al.*, 2014] by humans. One of the main reasons that users reject the use of prosthetic devices is the functional limitations of the limb. While state-of-the-art prosthetic limbs can perform complex functions and movements, control of this functionality by humans is still limited [Biddiss *et al.*, 2007]. New methods must be developed to help humans control complex robotic devices that are directly connected to them. Furthermore, such methods should incorporate some form of ongoing learning so that the device can adapt to the human who is wearing it.

Myoelectric prostheses are a class of modern robotic prosthesis which monitor electrical signals produced by muscle tissue in the patient's residual limb and use these signals to control the movement of a multiple-actuator robotic appendage [Parker et al., 2006]. Myoelectric control aims to remove functionality barriers for patients, but can be a challenge for new amputees; the transition to a powered prosthesis often requires extensive training and repeated calibration of the limb by clinicians. This difficulty is partly due to the user's inability to provide clear electromyographic (EMG) control signals, and partly due to control challenges in interpreting these complex multi-dimensional signals to guide robotic movements [Castellini *et al.*, 2014].

Another major limitation in most of the current prosthetic control schemes is the lack of adaptation over time. The need for adaptation may stem from changes in the patient physical and mental states, intents, and/or usage [Sensinger *et al.,* 2009]. Changing learned control policies requires expert knowledge of a patient's physiology and prosthetic hardware. Most users can not adaptively improve the control of their robotic limb independently, outside of the clinic.

A significant amount of past work has been done to address these challenges by incorporating machine learning into prosthetic control systems [Castellini *et al.*, 2014, Parker et al., 2006]. Examples include offline, supervised learning methods of dimensionality reduction [Englehart *et al.,* 2003, Artemiadis *et al.,* 2010], artificial neural networks, and support vector machines [Oskoei and Hu, 2008], as well as unsupervised [Sensinger *et al.,* 2009], and semi-supervised [Nishikawa *et al.,* 2001] techniques. A recent review by Castellini et al. provides a good overview of the state-of-the-art myoelectric prosthetic control research [Castellini *et al.*, 2014]. Adaptive, real-time approaches to addressing challenges in myoelectric control have also been proposed and shown to work in simulation, and in preliminary experiments with able-bodied subjects and subjects with amputations [Edwards *et al.*, 2015]. As well, previous work describes how these advances in prosthetic device control can be viewed as increasing communicative capital between two goal-seeking agents, the prosthetic and the human wearing it [Pilarski et al., 2015]. One promising adaptive method pairs human training of controllers with continuous-action actor-critic reinforcement learning (ACRL) [Degris et al., 2012, Pilarski *et al.,* 2011 and 2013]. We expand this ACRL approach in the present paper.

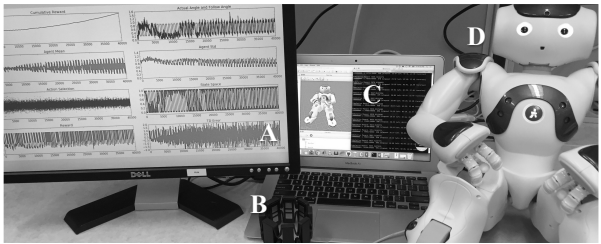

Figure 1. Configration with A) results, B) *Myo*, C) Simulation/Learning/Feedback System, and D) *Nao*.

Reinforcement learning (RL) was introduced as a theory of how humans and animals learn in response to positive and negative stimuli associated with actions towards a goal [Thorndike, 1898; Skinner, 1938]. In the early 1980's Sutton extended this concept to simulated learning agents and early work applied RL to difficult control problems [Sutton, 1984; Barto *et al.,* 1983]. RL is an approach for solving these control problems through interaction between a learning system and the environment, and provides a theoretical backing of convergence in well defined domains [Sutton and Barto, 1998].

RL methods, and specifically ACRL methods, are well suited to the prosthetic control learning task [Pilarski *et al.* 2011]. ACRL methods are parameter based, allowing linear, incremental computation of variables, which in turn allows fast updates even with very large problems. ACRL does not store past samples in memory, thus memory requirements remain constant through learning, allowing for long term learning and deployment on embedded hardware with limited resources. The methods can use temporally extended credit assignment, making it theoretically possible to learn quickly with sparse reward signals. When combined with function approximation, these methods can scale well to continuous-space real-world tasks [Thomas *et al.*, 2009]. Finally, ACRL provides an intuitive means for incorporation of human interaction in the form of scalar, goal-directed signals [Pilarski *et al.* 2011, Knox and Stone, 2015].

Similar past work has explored the applicability of incorporating ACRL methods in human-robot collaborative tasks. Peters and Schaal demonstrated ACRL learning of complex movement systems and motor primitives for humanoid robotics [Peters and Schaal, 2008], Izawa et al. showed that RL allowed a simulated biological two-joint arm to learn a reaching task [Izawa *et al.*, 2004], and Tamei and Shibata used a policy-based RL system which used EMG from an able bodied subject to control a robotic arm in a collaborative lifting task [Tamei and Shibata 2008].

Despite the body of past work, ACRL algorithms for the myoelectric control of prosthetics is still a largely unexplored domain. Most approaches in these problems rely on predefined reward signals based on knowledge of the task. This limits the ability for a user to adapt a control policy online based on a new task or goal. Further, it is not clear how well a user can provide both feedback and control signals, as this simultaneous interaction is not intuitive, and may demand a high cognitive burden for the human.

In this work we first review the necessary background as well as ongoing research in the field of incorporating real-time human-delivered feedback into a learning system. Then we explore a means by which the human and robot can collaborate interactively through control and feedback signals. We show how the human EMG signals define the robotic state space, and then explore the differences between using a conventional, task defined reward paradigms versus using human generated reward signal, and the combination of these reward sources. To demonstrate the challenges associated with simultaneous control of a robotic device and real-time direct human feedback delivering reward, we briefly describe the system configuration and user experience. Experiments in this paper provide new insight into the practicality of simultaneous human control and feedback signals in the training of robotic systems with many DoF.

## 2 Background

### 2.1 Reinforcement Learning

RL is a learning framework inspired by behaviorists [Thorndike, 1898; Skinner, 1938]. It describes how agents improve over time by taking actions in an environment with a goal of maximizing some reward signal [Sutton and Barto, 1998]. The control policy is iteratively improved by selecting the actions which maximize future reward signals accumulated by the agent. Commonly, RL problems are defined by Markov Decision Processes (MDPs) which are defined by the tuple: $(S, A, T, \gamma, R)$. $S$ defines the state, or the current observable environmental variables of the agent, $A$ defines the actions which the agent can take, $T$ is the transition probabilities between the current state *s* and the next state *s'*, given a specific action *a* was taken or more formally $T(s' | s, a)$, $\gamma$ defines the discount factor, or how much future reward is valued by the agent and takes a value between *[0,1]* where *1* values all possible future reward, and *0* values no future reward, *R* is a reward function based on state transitions.

The goal of an RL agent is to determine the correct action execution order, or behavior, to maximize the expected sum of future discounted reward, or *expected return*. This behavior is defined as the policy which dictates which action should be taken in a given state, or $\pi : S \rightarrow A$. RL approaches may be policy-based, searching the policy space directly, or value-based, estimating the value of possible actions in each state and then deriving a policy from these value estimations [Sutton and Barto, 1998]. The value, or quality, of a given state and action is defined by a Q-function, $Q : S \times A \rightarrow R$. Temporal difference algorithms, such as Q-learning [Watkins and Dayan, 1992], can approximate this function by iteratively updating an estimate temporal difference error.

The MDP may define either discrete or continuous state and action spaces. In continuous domains, the MDP contains an infinite set of states $s \in \mathbb{R}^{ns}$ and actions $a \in \mathbb{R}$, where *ns* is the number of dimensions of *S*. In this paper, we will denote $s_t, a_t, r_t$ as the state, action, and reward at time t; note that action $a_t$ may affect the reward accumulated on future time steps ($r_{t+1}, t > 0$) but not the current $r_t$.

## 2.2 The Continuous Actor-Critic Algorithm

In this paper we will focus on actor-critic algorithms, a subset of policy gradient based algorithms. In these algorithms, the control policy $\pi(a|s)$ is a function which defines the probability with which the system will select an action *a* in a state *s*. $\pi$ is characterized by a vector of parameters $\mathbf{w} \in \mathbb{R}^n$, we assume that for any state-action pair $\pi(a|s)$ is differentiable in **w**. The goal of policy-based methods is to find a policy $\pi$ which maximizes *expected return*. Policy-based methods update the policy parameter vector **w** in the direction of the gradient of the return with respect to **w** [Williams, 1992]. This gradient can be estimated from samples of interaction between the learning agent and the environment $(s_t, a_t, s_{t+1}, r_{t+1})$ [Sutton *et al.*, 1999]. This gradient estimation technique can have high variance, and thus require a large number of samples to converge.

Actor-critic (AC) methods aim to reduce this variance by using two learning systems: an *actor* and a *critic*. The actor shapes the policy $\pi$ and selects the actions, and the critic predicts the *expected return* while following the policy $\pi$. The critic represents the value function of the current policy $\pi$ in the state *s*. While this value function is not known, it can be estimated using temporal difference learning, this estimate is then used to compute the change in the parameter vector, or $\Delta \mathbf{w}$. Because *S* is continuous in the variables defining it, a standard function approximation technique known as tile coding is often used to transform the state s into a high-dimensional binary feature vector $x(s)$. This discretizes the space, and allows for generalization in the learned policies [Sutton and Barto, 1998]. Details in this paper are limited to those relevant to the experimental definition and results. More comprehensive details on ACRL algorithms are given by [Sutton and Barto, 1998] and [Peters and Schaal, 2008].

## 2.3 Incorporating Human Feedback

There are a variety of ways to incorporate human knowledge into a learning system, prior to, or during learning. Learning from demonstration, reward shaping, and inverse RL are all areas of active research closely connected with the work in this paper, a comprehensive review of those methods can be found elsewhere [Thomaz and Breazeal, 2008; Chernova and Tomaz 2014].

Knox and Stone introduced the *Interactive Shaping Problem* or how the feedback from the human can be best incorporated into the learning agent in a sequential decision making process [Knox et al. 2010]. Briefly, it is summarized as follows: given an agent is acting in an environment defined by an MDP, and a human who is observing the action performance and providing feedback to the agent, how can the agent learn the best possible task policy as measured by task performance or cumulative human feedback, given the information contained in the human feedback [Knox and Stone, 2009]. Further, Knox and Stone have delineated some of the biggest confounding issues in the Shaping Problem, namely the positive circuits problem and the credit assignment problem [Knox and Stone, 2012; Knox and Stone, 2015]. Previous work shows one method to address the credit assignment problem by using a linear model of credit assignment, which allows feedback to be applied to past time steps. The history window to which credit can be applied can be varied with a meta-parameter [Knox and Stone, 2009; Vien and Ertel, 2013]. Vien and Ertel also showed that the human feedback model can be generalized to address the problems associated with periods of noisy, or inconsistent, human feedback [Vien and Ertel, 2013]. Recent advancements in modelling human feedback with a Bayesian approach have improved on the work of Knox and Stone in discrete environments [Loftin *et al.*, 2015].

In this paper, we focus on incorporating direct human feedback into the MDP. This work is most similar to [Pilarski *et al.*, 2011 and 2013], where ACRL was used in a continuous robotic task. Those works used a variety of reward schemes, some of which depend on some knowledge of the problem domain. While more advanced, task specific reward schemes may help the agent converge faster than constant positive or negative rewards, this knowledge may not be available during learning, and/or the task definition may change.

## 3 Methods

### 3.1 Aldebaran Nao and Myo EMG Data

The experimental set up is shown in Figure 1. It is composed of the Aldebaran *Nao* robotic platform (Aldebaran Robotics), the wireless *Myo* EMG armband (Thalmic Labs), and a Mac Book Air (Apple, 2.2 GHz Intel Core i7, 8GB RAM) for human feedback and running the learning agent.

The Aldebaran Nao was chosen as a cost-effective physical test platform to experiment with algorithm development. Specifically, the upper arms were chosen to mimic some of the functionality of high DoF commercial myoelectric prostheses. Shown in Figure 1, the Nao has 25 DoF, the arm on the Nao has 6 DoF: shoulder pitch and roll, elbow yaw and roll, wrist yaw, and hand open/close. While the size of the Nao arm is less than that of commercial prosthetic devices, it provides a good test-bed for algorithmic development.

Experimentation took place on the physical device as well as on the accompanying simulation software for the Aldebaran Nao. This simulation software shares the dimensions and kinematic model of the physical device. On each time step, an angular change is sent to the joint which is autonomously controlled by the learning algorithm.

The EMG signals used in learning were recorded with the wireless *Myo* EMG armband. This mimics conventional control of some commercial myoelectric prostheses and the *Myo* has been used to supplement control signals in recent work from Johns Hopkins University. EMG data was sampled from all the electrodes, and a single control signal was computed by rectifying the signal on all the channels, then computing the mean of the forearm flexor muscle sensors and subtracting the mean of the forearm extensor muscle sensors, the output of this computation was a clean two-phase EMG-based control signal, $s_{EMG}$. These locations were selected to maximize the signal strength and to simplify the control signal. The computed EMG signal was time averaged as follows: $s_{EMG}(t+1) = (1-\tau) * s_{EMG}(t) + \tau * |s_{raw}(t)|$; where $s_{EMG}(0) = 0$, and the time constant which defines the past averaging was set to $\tau = 0.05$. While this signal preprocessing is theoretically non-essential in finding a successful policy, the dimensionality reduction of the state space decreased memory and processing requirements and allowed for faster policy convergence.

### 3.1 The Learning Algorithm

The continuous actor-critic algorithm used in this paper is a slightly modified version of an ACRL algorithm described previously [Pilarski et al. 2011; Pilarski et al., 2013]. The algorithm is detailed in Algorithm 1. The actor selects a new action a from a normal distribution $N(\mu, \sigma^2)$ defined by mean $\mu = w_\mu^T x(s)$ and standard deviation $\sigma = \exp(w_\sigma^T x(s))$. Actions in this work are defined as continuous angular increments or decrements. $w_\mu$ and $w_\sigma$ are parameter vectors for the mean and standard deviation of the actor system. The system selects and executes an action *a*, which transitions the agent into new state *s'* and generates a reward *r*, then the critic, which is defined by a parameter vector **v**, computes a *TD-error* from *r*, and the current estimates of the value of the old state *s* and the new state, *s'*. Each learning system (actor and critic) is defined by alpha, or scalar step-size parameters.

A major consideration in the RL framework is how to assign reward to past state-action pairs which may have contributed to the current state, this is known as the credit assignment problem. Eligibility traces are a common mechanism to help address this issue. Eligibility traces may be either accumulating or replacing, and are defined by traces on the states and trace decay rates, $\lambda$ [Sutton and Barto, 1998]. Replacing traces for the critic and accumulating traces for the actor are used to accelerate learning [Pilarski et al., 2013]. Often, more than one action is required by the agent, in these situations one set of actor parameters may be maintained for each action [Pilarski et al., 2011 and 2013, Tamei and Shibata 2009].

## 4 Experiments

This work explores the incorporation of simultaneous control and direct human feedback into a RL agent running on an Aldebaran *Nao*. The goal of the task was to output joint velocity commands to the Nao's left arm to track a two-phase, periodic, target angle trajectory, demonstrated with the right arm, well within the maximum joint angle range.

In learning an optimal control policy in a simple motion tracking sequence, 4 experimental reward conditions were considered: 1) fixed goal-based reward where $r_{fixed} = 1$ if the joint is within a deviation threshold, $\Delta\theta_{max}$, of the target joint angle, $\theta_t$, else $r_{fixed} = -0.5$, 2) relative goal-based reward where $r_{relative} = 1$ if the joint is within $\Delta\theta_{max}$ of $\theta_t$, else $r_{relative} = -|\theta - \theta_t|$, 3) strictly human-delivered reward where the human could deliver a reward of $r_h = 1 \text{ or } -0.5$ by pressing corresponding positive or negative keys, and 4) learning from the combination of fixed goal-based and human reward conditions, $r_{total} = r_{fixed} + r_h$.

The action space of the learning system consisted of a single continuous angular displacement value for the Left Elbow Roll of the Aldebaran *Nao*. This angular displacement value was clipped in the range [-0.05, 0.05] to ensure smooth motion and exploration of the state space. As well, actions which would take the joint outside of the *Nao*'s allowable range were clipped to their allowable values. Actions were selected and performed on each time step as shown in Algorithm 1. A continuous state space consisting of $s = <\theta_{er}, s_{EMG}>$ was used, where $\theta_{er}$ indicated the left elbow roll joint angle, and $s_{EMG}$ was the EMG control signal computed as described in Section 3.1. All the components in *s* were normalized to the range *[0, 1]* according to their maximum and minimum possible values prior to use in function approximation [Pilarski et al., 2011 and 2013].

The continuous state space was function approximated using tile coding to allow for discretization and generalization of the state space [Pilarski et al., 2011 and 2013]. Tile coding was used to construct the state approximation vector *x(s)* used in learning. To capture a variety of generalization levels, *x(s)* was a concatenation which combined $N_T = 5$ offset tilings of *s* (which is 2-dimensional), at *3* different resolution levels, $N_R = [3,5,8]$, along with a single active baseline unit. This resulted in a single binary feature vector consisting of $(5 * 3^2) + (5 * 5^2) + (5 * 8^2) + 1 = 491$ features, exactly $m = 16$ features in *x(s)* were active at a given time, one for each tiling at each resolution and one for the active baseline feature. The learning parameters were set as follows: $\alpha_v = 0.01/m$, $\alpha_\mu = 0.005/m$, $\alpha_\sigma = \alpha_\mu$,

---

**Algorithm 1** Continuous Actor-Critic RL Algorithm

1: **initialize:** $\mathbf{w}_\mu, \mathbf{w}_\sigma, \mathbf{v}, \mathbf{e}_\mu, \mathbf{e}_\sigma, \mathbf{e}_\mathbf{v}, s$
2: **repeat:**
3:      $\mu \leftarrow \mathbf{w}_\mu^T \mathbf{x}(s)$
4:      $\sigma \leftarrow \exp[\mathbf{w}_\sigma^T \mathbf{x}(s)]$
5:      $a \leftarrow \mathcal{N}(\mu, \sigma^2)$
6:      **take action** $a$, **observe** $r, s'$
7:      $\delta \leftarrow r + \gamma \mathbf{v}^T \mathbf{x}(s') - \mathbf{v}^T \mathbf{x}(s)$
8:      $\mathbf{e}_\mathbf{v} \leftarrow min[1, \lambda_\mathbf{v} \gamma \mathbf{e}_\mathbf{v} + \mathbf{x}(s)]$
9:      $\mathbf{v} \leftarrow \mathbf{v} + \alpha_\mathbf{v} \delta \mathbf{e}_\mathbf{v}$
10:     $\mathbf{e}_\mu \leftarrow \lambda_\mathbf{w} \mathbf{e}_\mu + (a - \mu)\mathbf{x}(s)$
11:     $\mathbf{w}_\mu \leftarrow \mathbf{w}_\mu + \alpha_\mu \delta \mathbf{e}_\mu$
12:     $\mathbf{e}_\sigma \leftarrow \lambda_\mathbf{w} \mathbf{e}_\sigma + [(a - \mu)^2 - \sigma^2]\mathbf{x}(s)$
13:     $\mathbf{w}_\sigma \leftarrow \mathbf{w}_\sigma + \alpha_\sigma \delta \mathbf{e}_\sigma$
14:     $s \leftarrow s'$

$\gamma = 0.9$, $\lambda_w = 0.3$, $\lambda_v = 0.7$, joint angles were limited by manufacturer specifications at $\theta \in [0.0349, 1.5446]$. Weight vectors $e_v$, $e_\mu$, $e_\sigma$, $v$, $w_\mu$ and $w_\sigma$ were initialized to 0 and standard deviation was bounded by $\sigma \geq 0.01$. Parameters for the experimentation were set as follows: maximum number of time steps = 40000, frequency of iteration was ~33 Hz or , and angular deviation threshold was set to $\Delta\theta_{max} = 0.1$, actions were selected and performed on every time step.

The ACRL system was trained online, with EMG control signals being read directly from the user. Learning updates and action choice occurred at ~33Hz or every ~30 ms. Total training time steps were held constant for the experiments. Performance was measured by taking the average mean absolute angular error from 1) all the data, 2) the last 10k steps, 3) the last 5k steps. This was done to compare the experimental results after some learning was completed; this helped to reduce the noise intrinsic in early learning.

With this experimental set up we were able to test our four experimental conditions on the physical robot, with live EMG fed into the learning system, and we were able to test the performance of the learning using a simulated robotic agent, and a simulated EMG source. These simulations allowed us to rapidly test the learning over a large number of repetitions. Figure 3 shows the comparison between the physical and simulated robot and real and simulated EMG.

For the human EMG experiments the subject (1 healthy male, 28 yr.) gave informed consent to participate, and the trial was approved by the human research ethics board at the University of Alberta. The subject was directed to perform a repeated task; they were instructed to follow the movement of the robotic arm in reaching and retracting by flexing or relaxing their wrist. When the EMG signal was simulated it was derived from a noisy estimation of the target signal. As previously described by [Pilarski et al., 2011], to provide a human training signal (experimental conditions 3 and 4) across the fast learning algorithm, the reward signal on each time step following delivery was set to a decayed trace of the reward of the previous step, or: $r_{t+1} = 0.01\, r_t + r_h$. This allowed human delivered feedback, which occurs on the second time scale to be smeared across actions occurring over the millisecond time scale.

## 5 Results

Our results demonstrate a potential benefit to be had by introducing human feedback into the robotic learning system. The inclusion of human shaping signals was shown to improve performance over strictly environmentally derived reward, but increased the level of cognitive attention required of the user during training.

Figure 2 shows the results of a single, representative, 40k time step learning experiment for combined human and fixed-goal based reward. It shows the total accumulated reward, the actual target angle and learning system actuated angle, as well as the learning agent mean and standard deviation (as calculated in Algorithm (1). On the *Actual and Follow Angle* plot the target angle is light gray, the learning agent controlled joint is dark gray and the angular distance error threshold is shown dotted around the target angle. The agent learns a near optimal policy after ~15k time steps, after which it accumulates reward consistently. The agent mean and standard deviation plots show multiple phases learned by the agent; these are the repeating locations on the target trajectory learned as the agent gains certainty in optimal actions to select in each phase.

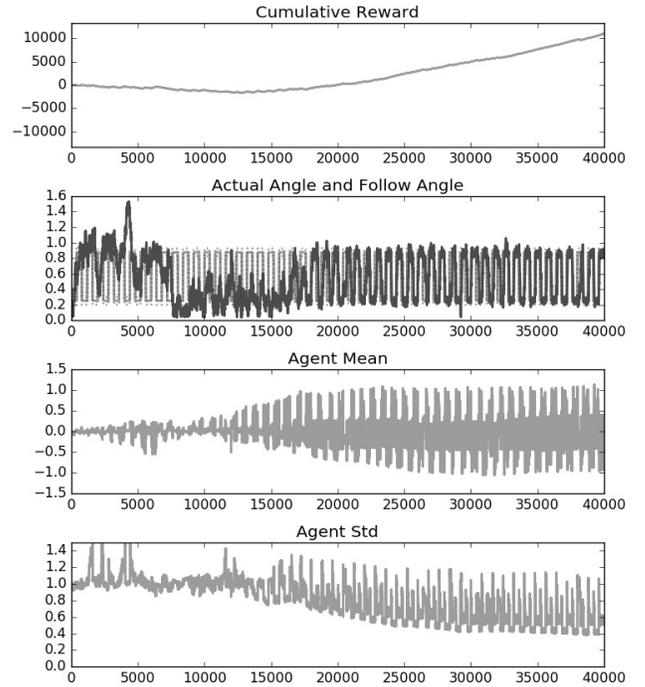

Figure 2. Representative results from 40k learning steps with human reward combined with fixed-goal based reward.

Figure 3 compares learning performance on the simulation and physical robotic and EMG configurations. It shows the mean angle error (MAE) over varying numbers of times steps. MAE is the difference between the target and the learning system controlled joint angle, averaged over a) all time steps, b) the last 10k time steps and c) the last 5k time steps. For several conditions, multiple trials were run, as denoted by (n = number of trials) in the legend. In these cases, mean and standard deviation over the number of trials is reported. In the legend, when *Nao* or *Myo* is listed, it implies that this performance was measured on the physical robot or Myo system as opposed to simulation.

Figure 3 shows MAE over all time steps, the last 10k, and last 5k time steps for 1) constant reward, 2) relative reward, 3) strictly human reward, and 4) constant + human reward sources. The graph also has comparitive data for physical and simulation robotic systems and EMG. When *Myo* or *Nao* is listed (e.g. C,D,G,H), it implies the physical system was used, otherwise, are from a simulation robot and/or simulated EMG data. Figure 3 shows a close correlation between real and simulated EMG signal experimental conditions, and an important distinction between performance when learning on a simulated versus a physical robot. This result is to be expected; while the EMG control signal is not noisy and is simple to simulate, the difference between a simulated versus physical robot is significant. As can be seen in Figure 3, the performance of relative error reward is superior to constant error reward learning in the simulated robotics. This provides an indication that our learning algorithm is utilizing the information contained within the reward signal to refine its policy and action selection. The results also indicate that strictly human reward based training performs the poorest over the given numbers of time steps. This is most likely due to the noise inherent in human delivered reward, as well as the potential errors in the human reward decay trace calculation. Figure 3 shows there may be benefit to incorporating human feedback into the learning system, potentially limited by the richness of the environmental reward. The MAE for *Constant Reward* on the simulated *Nao* with simulated EMG data is greater than the value for when human feedback is included. The MAE for *Constant Reward with Human Feedback* on the real *Myo* and *Nao* falls within the standard deviation of the simulated systems. This may be expected; while learning on physical robots may hinder, human feedback may improve learning.

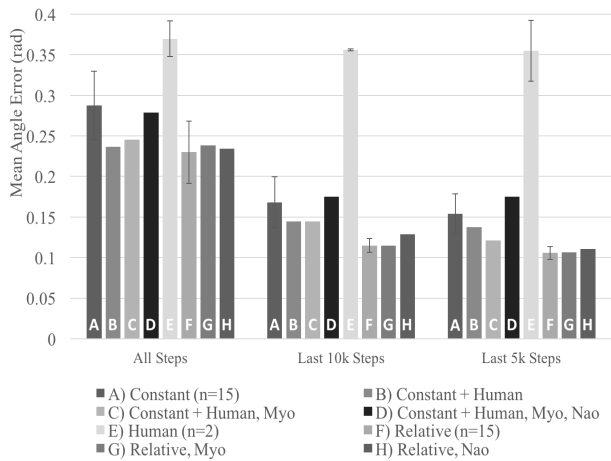

Figure 3. Comparitive bar graph showing the aggregate data over the multiple trials of each experimental condition.

## 6 Discussion

This work explores combinations of human and environmental feedback in simulated and physical robotic systems with simulated and real EMG control signals. Simulation and physical experimentation show agreement, but some performance decrease may be expected with physical robotic systems. This performance decrease may be due heat, inertia, and/or mechanical jitter. While there are benefits to development and testing on a simulated robot, including safety, controlled repeatability and batch processing, additional experimentation on physical robots is necessary to support human-feedback derived improvement.

To speed up learning step size, $\gamma$, EMG and human reward decay parameters, and tile coding parameters were selected to maximize the reward accumulated on the simulated *Myo* and *Nao*. These parameters may not be optimal for the physical *Nao* configuration, and could potentially be adapted online in the future [Sherstov and Stone, 2005].

The results indicate the poorest learning performance in strictly human feedback based learning. There are challenges associated with simultaneously providing control and feedback signals. The cognitive burden is demanding, and it can be difficult to maintain attention and focus when providing both inputs. As well, providing clear, consistent feedback to the robot is difficult [Chernova and Tomaz 2014]. Shifts in concentration, distractions, or confusion on the intention of positive and negative feedback can result in noisy feedback signals. The research area exploring how to model and deliver consistent human feedback is rich with methods, such as reward shaping [Brys, 2015], TAMER [Knox and Stone, 2009], iSABL [Loftin *et al.*, 2015], and Actor-critic and Tile-coding TAMER [Vien and Ertel, 2013]. These algorithms generalize and model human feedback, and claim to extract additional shaping value based on predictions of the human proving the feedback.

The combination of the MDP and human-delivered reward signals may modify the agent's task definition. If human feedback is inconsistent, the changing task definition may cause instability in the system. Potential-based reward shaping may address this and should be explored [Harutyunyan et al., 2015].

In this experiment only single joint control was learned; it is possible to learn multiple joint policies concurrently. In these paradigms, it is often valuable to break down the task into regions of solvability to decrease the total time needed to learn complex composite behaviors. This scaffold approach has been shown to decrease convergence times, and make interactive human training easier and is a rich area for future research [Pilarski et al., 2011, Sanger, 1994].

The main results reported here is the MAE overall and late learning. Additional performance metrics to be explored include convergence time, number of human interactions, and more qualitative measures including cognitive burden.

## 7 Conclusions

This paper contributes a first set of results on incorporating simultaneous human control and feedback signals in the training of a semi-autonomous robotic agent. Our results demonstrate performance increases by incorporating human-feedback into existing algorithms, and show that human interaction may improve performance in complex robotic tasks. This work therefore provides a new viewpoint on the human training of a robotic system tightly coupled to a user's body, and suggests that seamless interaction with complex hardware may demand a shift of autonomy from the human to the machine. Future work will explore methods to reduce cognitive burden on the human and to model, generalize, and deliver consistent human feedback.

### Acknowledgments

This work is supported by National Sciences & Engineering Research Council of Canada, Alberta Innovates – Technology Futures, and the Alberta Innovates Centre for Machine Learning.

# References

[Artemiais and Kyriakopolous, 2010] Artemiadis PK, Kyriakopoulos KJ. EMG-based control of a robot arm using low-dimensional embeddings. *IEEE Trans. Robotics*. 2010. 26(2):393-8.

[Barto et al., 1983] Barto AG, Sutton RS, Anderson CW. Neuronlike adaptive elements that can solve difficult learning control problems. *IEEE Trans. Systems, Man & Cybernetics*. 1983. (5):834-46.

[Biddiss and Chau, 2007] Biddiss EA, Chau TT. Upper limb prosthesis use and abandonment: a survey of the last 25 years. *J Prosthet Orthot Intl*. 2007. 31(3):236-57.

[Brys et al., 2015] Brys, T., Harutyunyan, A., Suay, H.B., Chernova, S., Taylor, M.E. and Nowé, A., 2015, June. Reinforcement learning from demonstration through shaping. *In Proceedings of the International Joint Conference on Artificial Intelligence*.

[Castellini et al., 2014] Castellini C, Artemiadis P, Wininger M, Ajoudani A, Alimusaj M, Bicchi A, Caputo B, Craelius W, Dosen S, Englehart K, Farina D. *Proc. Of 1$^{st}$ Workshop on Peripheral Machine Interfaces*. 2014.

[Chernova and Tomaz 2014] Chernova S, Thomaz AL. Robot learning from human teachers. *Syn. Lecs. on AI and ML*. 2014. 8(3):1-21.

[Degris et al., 2012] Degris T, Pilarski PM, Sutton RS. Model-free reinforcement learning with continuous action in practice. *American Control Conference*. 2012 Jun 27 (pp. 2177-2182). IEEE.

[Edwards *et al.*, 2015] Edwards AL, Dawson MR, Hebert JS, Sherstan C, Sutton RS, Chan KM, Pilarski PM. Application of real-time machine learning to myoelectric prosthesis control: A case series in adaptive switching. *J Prosthet Orthot Int*. 2015.

[Englehart and Hudgins, 2003] Englehart K, Hudgins B. A robust, real-time control scheme for multifunction myoelectric control. *IEEE Trans. Biomed Eng*. 2003. 50(7):848-54.

[Harutyunyan et al., 2015] Harutyunyan, A., Brys, T., Vrancx, P. and Nowé, A., 2015, May. Shaping mario with human advice. In *Proceedings of the 2015 Int. Conference on Autonomous Agents and Multiagent Systems* (pp. 1913-1914). International Foundation for Autonomous Agents and Multiagent Systems.

[Izawa and Ito, 2004] Izawa J, Kondo T, Ito K. Biological arm motion through reinforcement learning. *Biological cybernetics*. 2004. 91(1):10-22.

[Knox and Stone, 2009] Knox WB, Stone P. Interactively shaping agents via human reinforcement: The TAMER framework. *In Proc of 5$^{th}$ Intl. Conf. on Knowledge Capture*. 2009 (9-16). ACM.

[Knox and Stone, 2012] Knox WB, Stone P. Reinforcement learning from human reward: Discounting in episodic tasks. *IEEE RO-MAN*. 2012. (pp. 878-885).

[Knox and Stone, 2015] Knox WB, Stone P. Framing reinforcement learning from human reward: Reward positivity, temporal discounting, episodicity, and performance. *Artificial Intelligence*. 2015. 225:24-50.

[Loftin et al., 2015] Loftin R, Peng B, MacGlashan J, Littman ML, Taylor ME, Huang J, Roberts DL. Learning behaviors via human-delivered discrete feedback: modeling implicit feedback strategies to speed up learning. *AAMAS*. 2016. 30(1):30-59.

[Nisikawa et al., 2001] Nishikawa D, Yu W, Yokoi H, Kakazu Y. On-line learning method for EMG prosthetic hand control. *Elec. and Comm Japan, III*. 2001. 84(10):35-46.

[Oskoei and Hu, 2008] Oskoei MA, Hu H. Support vector machine-based classification scheme for myoelectric control applied to upper limb. *IEEE Trans. Biomed Eng*. 2008. 55(8):1956-65.

[Parker et al, 2006] Parker P, Englehart K, Hudgins B. Myoelectric signal processing for control of powered limb prostheses. *J Electromyogr Kinesiol*. 2006. 16(6):541-8.

[Peters and Schaal, 2008] Peters J, Schaal S. Natural actor-critic. *Neurocomputing*. 2008. 71(7):1180-90.

[Peters et al., 2003] Peters J, Vijayakumar S, Schaal S. Reinforcement learning for humanoid robotics. *3$^{rd}$ IEEE-RAS Intl. Conf. on Humanoid Robots*. 2003 (pp. 1-20).

[Pilarski et al., 2011] Pilarski PM, Dawson MR, Degris T, Fahimi F, Carey JP, Sutton RS. Online human training of a myoelectric prosthesis controller via actor-critic reinforcement learning. *IEEE Intl. Conf. on Rehabilitation Robotics*. 2011. (pp. 1-7). IEEE.

[Pilarski et al., 2013] Pilarski PM, Dick TB, Sutton RS. Real-time prediction learning for the simultaneous actuation of multiple prosthetic joints. *IEEE Intl. Conf. on Rehabilitation Robotics*. 2013. (pp. 1-8). IEEE.

[Pilarski et al., 2015] Pilarski PM, Sutton RS, Mathewson KW. Prosthetic Devices as Goal-Seeking Agents. 2$^{nd}$ *Workshop on Present and Future of Non-Invasive Peripheral-Nervous-System Machine Interfaces*, Singapore, 2015.

[Sanger, 1994] Sanger TD. Neural network learning control of robot manipulators using gradually increasing task difficulty. *IEEE Trans. Robotics and Automation*. 1994. 10(3):323-33.

[Sensinger et al., 2009] Sensinger JW, Lock BA, Kuiken TA. Adaptive pattern recognition of myoelectric signals: exploration of conceptual framework and practical algorithms. *IEEE Trans. Neural Sys. and Rehab. Eng*. 2009. 17(3):270-8.

[Skinner, 1938] Skinner BF. The behavior of organisms: an experimental analysis. *Appleton-Century, Oxford*. 1938.

[Sutton, 1984] Sutton RS. Temporal credit assignment in reinforcement learning. *Doctoral Dissertation*. 1984.

[Sutton and Barto, 1998] Sutton RS, Barto AG. Reinforcement learning: An introduction. *MIT Press*. 1998.

[Sutton et al., 1999] Sutton RS, McAllester DA, Singh SP, Mansour Y. Policy Gradient Methods for Reinforcement Learning with Function Approximation. *NIPS*. 1999 (Vol. 99, pp. 1057-1063).

[Tamei and Shibata, 2009] Tamei T, Shibata T. Policy gradient learning of cooperative interaction with a robot using user's biological signals. *Adv. In Neuro-Information Processing*. 2008 (pp. 1029-1037). Springer Berlin Heidelberg.

[Thomaz and Breazeal, 2008] Thomaz AL, Breazeal C. Teachable robots: Understanding human teaching behavior to build more effective robot learners. *Artificial Intelligence*. 2008. 172(6):716-37.

[Thomas et al., 2009] Thomas P, Branicky M, van den Bogert A, Jagodnik K. Application of the actor-critic architecture to functional electrical stimulation control of a human arm. Conf. *Proc. Innovative Applications of AI*. 2009 (p. 165).

[Thorndike, 1898] Thorndike EL. Animal Intelligence. *Nature*. 1898. 58:390.

[Vien and Ertel, 2013] Vien NA, Ertel W, Chung TC. Learning via human feedback in continuous state and action spaces. *Applied Intelligence*. 2013. 39(2):267-78.

[Watkins and Dayan, 1992] Watkins CJ, Dayan P. Q-learning. *Machine Learning*. 1992. 8(3-4):279-92.

[Williams, 1992] Williams RJ. Simple statistical gradient-following algorithms for connectionist reinforcement learning. *Machine Learning*. 1992. 8(3-4):229-56.